\documentclass[twocolumn,epjc3]{svjour3}  
\smartqed
\usepackage[utf8,latin1]{inputenc}
\newcommand{\qm}[1]{``#1''}
\usepackage{amsmath}
\usepackage{amssymb}
\usepackage{mathabx}
\usepackage{mathrsfs}  
\usepackage{cases}
\usepackage{colortbl}
\usepackage{dcolumn}
\usepackage{xcolor}
\usepackage{accents}
\usepackage{dcolumn}
\usepackage{xcolor,colortbl}
\usepackage{multirow}
\usepackage{stackengine}
\usepackage{bm}
\usepackage{bbm}
\usepackage{hyperref}
\hypersetup{colorlinks, linkcolor={red},citecolor={blue},urlcolor={blue}}  
\usepackage[square,numbers,sort&compress]{natbib}
\usepackage{stackengine,scalerel}

\newcommand{\lc}[1]{\accentset{\circ}{#1}}

\newcommand{\pa}{\partial}
\newcommand{\del}[1]{\delta_g #1}
\newcommand{\pg}[1]{\frac{\pa #1}{\pa g^{\mu\nu}}}

\newcommand{\dd}{{\rm d}}
\RequirePackage{graphicx}
\journalname{Eur. Phys. J. C}

\RequirePackage{graphicx}
\journalname{Eur. Phys. J. C}

\begin{document}

\title{The role of the boundary term in  $f(Q,B)$ symmetric teleparallel gravity}
\titlerunning{The role of the boundary term in  $f(Q,B)$ symmetric teleparallel gravity}      

\author{Salvatore Capozziello\thanksref{e1,addr1,addr2,addr3}
\and
Vittorio De Falco\thanksref{e2,addr2,addr3}
\and
Carmen Ferrara\thanksref{e3,addr2,addr3}}

\thankstext{e1}{e-mail: capozziello@na.infn.it}
\thankstext{e2}{e-mail: vittorio.defalco-ssm@unina.it}
\thankstext{e3}{e-mail: carmen.ferrara-ssm@unina.it}

\authorrunning{Capozziello et al. (2023)}

\institute{Dipartimento di Fisica \qm{E. Pancini}, Universit\'a di Napoli \qm{Federico II}, Complesso Universitario di Monte S. Angelo, Via Cintia Edificio 6, I-80126 Napoli, Italy \label{addr1}
\and
Scuola Superiore Meridionale, Largo San Marcellino 10, 80138 Napoli, Italy \label{addr2}
\and
Istituto Nazionale di Fisica Nucleare, Sezione di Napoli, Complesso Universitario di Monte S. Angelo, Via Cintia Edificio 6, 80126 Napoli, Italy  \label{addr3}}

\date{Received: \today / Accepted: }

\maketitle

\begin{abstract}
In the framework of metric-affine gravity, we consider  the role of the boundary term in   Symmetric Teleparallel Gravity assuming $f(Q,B)$ models  where $f$ is a smooth function of the non-metricity scalar $Q$ and the  related boundary term $B$. Starting from  a variational approach, we derive the  field equations and compare them with respect to those of  $f(Q)$ gravity in the limit of $B\to0$. It is possible to show that    $f(Q,B)=f(Q-B)$ models  are dynamically equivalent to $f(R)$ gravity  as in the case of teleparallel $f(\tilde{B}-T)$ gravity  (where $B\neq \tilde{B}$).  Furtherrmore, conservation laws are derived.  In this perspective, considering boundary terms in $ f(Q)$ gravity represents the last ingredient towards  the Extended Geometric Trinity of Gravity, where $f(R)$, $f(T,\tilde{B})$, and $f(Q,B)$ can be dealt with under the same standard. We also compare and discuss about the Gibbons-Hawking-York boundary term of General Relativity and the boundary term $B$ in $f(Q,B)$ gravity.
\end{abstract}

\section{Introduction}
\label{sec:intro}
General Relativity (GR) is the best and well-tested theory of gravity so far available. However, it cannot be the final model of gravitational interaction, because of some fundamental reasons related to its infrared (IR) and ultraviolet (UV) behaviors. At astrophysical and  cosmological scales,  we need  mechanisms capable of  explaining the clustering of structures and the accelerated expansion\footnote{Difficulties in detecting new fundamental particles  to explain the dark sector point out  that alternative approaches could be pursued in order to better match the observed phenomenology.} \cite{Carroll2001,Rocco2019,Planckcollaboration2020}. Furthermore, at microscopic scales, GR results to be not renormalizable \cite{tHooft1974,Goroff1985}. The fact that a coherent quantum gravity theory  does not yet exist \cite{Hooft1974,Gleiser2005} means that some extensions or modifications of GR are needed. To address the aforementioned issues, the current most-followed strategy consists of formulating gravitational theories admitting  GR in some limits and including, in general, further degrees of freedom \cite{Clifton2006, Faraoni2010,Odintsov2011,Capozziello2011R,Oikonomou,Celia,Celia2,Lavinia}. All these theories come from  motivations related to the fundamental structure and principles of gravitational field (see Refs. \cite{CANTATA,Bohmer2021,Bohmer2023}, for comprehensive reviews).

In this debate, the validity of the Equivalence Principle at quantum scales, as well as related features like causal and geodesic structure play a main role. In particular, in view of dealing with gravity as a gauge theory, the metricity requirement is of paramount importance. Indeed, relaxing the hypothesis that metricity principle holds in any case can offer a possibility to achieve a fundamental theory overcoming the GR shortcomings. Among metric-affine theories, Symmetric\\ Teleparallel Gravity and its extensions are assuming  a prominent role in the discussion to build up a final theory of gravity. In this approach, the dynamics is described by the non-metricity scalar $Q$, derived from the non-metricity tensor 
$Q_{\alpha\mu\nu}=\nabla_\alpha g_{\mu\nu}$ (see e.g. \cite{Beltran2018,DAmbrosio2022}). Recently, the extension $f(Q)$, where $f$ is a smooth function of $Q$, has been  exploited to describe bouncing cosmology \cite{Bajardi2020}, late time accelerated expansion \cite{Rocco2022,Solanki2022}, and early time inflationary behavior \cite{Shokri}, as well as to constrain gravitational wave observations and to test GR \cite{Soudi2019,Dagostino2022}. There are also applications in high-energy astrophysics for black hole \cite{DAmbrosio2022, Mustafa} and wormhole  \cite{Parsaei2022, Sharma} solutions. Other fundamental developments are related to Noether symmetries \cite{Kostas} and  minisuperspace quantum cosmology \cite{Quantum} in $f(Q)$ gravity.

In particular, for $f(Q)=a\ Q+b$, with $a,b$ real constants, it is possible to retrieve the Symmetric Teleparallel Equivalent of GR (STEGR), where the Lagrangian of GR and STEGR differ only for a boundary term $B$ \cite{Capozziello2022}. In other words, the two mathematical frameworks produce exactly the same field equations, albeit they seem to be \emph{a-priori} completely unrelated theories. Beside them, there is also the Teleparallel Equivalent of GR (TEGR), based on the torsion scalar $T$ and the related boundary term $\tilde{B}$ (with $B\neq\tilde{B}$), which is another equivalent formulation of GR \cite{Pereira,Cai,Capozziello2022}. The aforementioned three theories, whose dynamics are encoded in the Ricci curvature scalar $R$ for GR, the torsion scalar $T$ for TEGR, and the non-metricity scalar $Q$ for STEGR, constitute the so-called \emph{Geometric Trinity of Gravity}  \cite{Tomi,Capozziello2022}.

The goal of this article aims at presenting the  $f(Q,B)$ gravity, configuring as an extension of $f(Q)$ gravity and, more in general, of STEGR. We analyse some features of this model and we discover that $f(Q,B)=f(Q-B)$ represents a dynamically equivalent formulation of $f(R)$ gravity (being an extension of GR), where $f$ is a smooth function of $Q,B$ in the former case and of $R$ in the latter occurence. The paper is organized as follows: in Sec. \ref{sec:f(Q)_gravity} we briefly recall the $f(Q)$ gravity and set out our notations.  Sec. \ref{sec:f(Q,B)_gravity} is devoted to  modified $f(Q,B)$ gravity model verifying the consistency of the obtained equations and introducing the concept of Extended Geometric Trinity of Gravity. A comparison between boundary terms appearing in GR and those belonging to $f(Q)$ theory is developed.  In Sec. \ref{sec:end}, we draw the conclusions.

\emph{Notations.}   The spacetime metric is  $g_{\mu\nu}=\eta_{AB} e^A_{\mu}e^B_{\nu}$ where $e^A_{\mu}$ are the tetrad fields on the tangent space with Minkowskian metric $\eta_{AB}$.  The determinant of the metric $g_{\mu \nu}$ is denoted by $g$ and $e=\sqrt{-g}$. Round (square) brackets around a pair of indices stands for the  symmetrization (antisymmetrization) procedure, i.e., $A_{(ij)}=A_{ij}+A_{ji}$ (respectively, $A_{[ij]}=A_{ij}-A_{ji}$).  All quantities with an over-circle denote objects framed in GR, like $\lc{\Gamma}^\lambda{}_{\mu\nu},\lc{\nabla}_\alpha$; whereas quantities without any marked symbols are framed in the Symmetric Teleparallel Gravity, like $\Gamma^\lambda{}_{\mu\nu},\nabla_\alpha$. The coupling constant in the metric field equations is $\chi=\frac{8\pi G}{c^4}$. We indicate the partial derivatives of $f$ with $f_X(X,Y)=\frac{\pa f}{\pa X}$, and $f_{XX}(X,Y)=\frac{\pa{}^2f}{\pa X^2}$. The same conventions hold also with respect to $Y$ or mixed derivatives with respect to $X$ and $Y$.

 \section{$f(Q)$ symmetric teleparallel gravity}
\label{sec:f(Q)_gravity}
A subclass of metric-affine geometries is represented by the Symmetric Teleparallel Gravity theories, characterized by vanishing curvature and torsion \cite{Capozziello2022}. The only surviving quantity is the non-metricity tensor
\begin{subequations}
\begin{align}
Q_{\alpha\mu\nu}&=\nabla_\alpha g_{\mu\nu}=\partial_\alpha g_{\mu\nu}-\Gamma^\lambda_{\alpha\mu}g_{\nu\lambda}-\Gamma^\lambda_{\alpha\nu}g_{\mu\lambda},\label{eq:nonm_ten}
\end{align}
\end{subequations}
expressing the failure of the metric compatibility when different from zero. In this framework, metric and affine connection are two independent geometrical objects. The former is deputed to define the \emph{casual structure}, whereas the latter  describes the \emph{geodesic structure}. 

Specifically, it is possible to take into account   the $f(Q)$ gravity, expressed in terms of the following action
\begin{equation} \label{eq:f(Q)_action}
S_Q=\int \dd^4x\ e \biggr{[}\frac{1}{2\chi}f(Q)+\mathcal{L}_m\biggr{]},
\end{equation}
where $Q$ is the \emph{non-metricity scalar} defined as 
\begin{equation}
Q=-\frac{Q_{\alpha\mu\nu}Q^{\alpha\mu\nu}}{4}+\frac{Q_{\alpha\mu\nu}Q^{\alpha\nu\mu}}{2}+\frac{Q_\alpha Q^\alpha}{4} -\frac{Q_\alpha \bar{Q}^\alpha}{2},
\end{equation}
where $Q_\alpha=Q_{\alpha\mu}{}^\mu$ and $\bar{Q}_\alpha=Q^\mu_{\ \mu\alpha}$. The non-metricity scalar can be also written as follows
\begin{equation} \label{eq:boundary_term}
\lc{R}=Q-B,\qquad B=\lc{\nabla}_\lambda\tilde{Q}^\lambda=\frac{1}{e}\partial_\lambda(e \tilde{Q}^\lambda),
\end{equation}
where $\tilde{Q}^\lambda=Q^\lambda-\bar{Q}^\lambda$, $\lc{R}$ is the GR Ricci scalar curvature, $\lc{\nabla}$ the GR covariant derivative, and $B$ the boundary term. 

If the action $S_Q$ is derived with respect to the metric (i.e., $\delta_g S_Q=0$), we obtain the \emph{metric field equations of second-order in $g_{\mu\nu}$}, namely $\mathcal{M}_{\mu\nu}=\chi \Theta_{\mu\nu}$ (see Appendix \ref{sec:f(Q)_gravity_VF}, for derivation and  details), with \cite{DAmbrosio2022}
\begin{subequations} 
\begin{align} 
\mathcal{M}_{\mu\nu}&=\frac{2}{e}\nabla_\alpha(eP^\alpha_{\ \mu\nu}f_Q)+\frac{1}{e}q_{\mu\nu}f_Q-\frac{1}{2}g_{\mu\nu}f, \label{eq:field-equations}\\
\Theta_{\mu\nu}&=-\frac{2}{e}\frac{\partial\mathcal{L}_m}{\partial g^{\mu\nu}}, \label{eq:stress-energy-tensor}
\end{align} 
\end{subequations} 
where 
\begin{subequations} 
\begin{align} 
P^\alpha_{\ \mu\nu}&=\frac{1}{2e}\frac{\partial(e Q)}{\partial Q_\alpha{}^{\mu\nu}}=\notag\\
&=-\frac{Q^\alpha_{\ \mu\nu}}{4}+\frac{Q_{(\mu}{}^\alpha_{\ \nu)}}{4}+\frac{g_{\mu\nu}\tilde{Q}^\alpha}{4}-\frac{\delta^\alpha_{\ (\mu}Q_{\nu)}}{8},\label{eq:P}\\
\frac{1}{e}q_{\mu\nu}&=\frac{1}{e}\frac{\partial(e Q)}{\partial g^{\mu\nu}}+\frac{1}{2}g_{\mu\nu}Q\notag\\
&=P_{\nu\rho\sigma}Q_\mu{}^{\rho\sigma}-2P_{\rho\sigma\mu}Q^{\rho\sigma}{}_\nu-\frac{1}{2}g_{\mu\nu}Q.\label{eq:qmunu}
\end{align} 
\end{subequations} 
$\mathcal{M}_{\mu\nu}$ can be recast in the GR-like form as follows\cite{DAmbrosio2022}
\begin{align} 
\mathcal{M}_{\mu\nu}&=f_Q\lc{G}_{\mu\nu}-\frac{1}{2}g_{\mu\nu}\biggr{[}f-f_QQ\biggr{]}+2P^\alpha_{\ \mu\nu}\pa_\alpha f_{Q},\label{eq:FE}
\end{align}
where $\lc{G}_{\mu\nu}$ is the GR Einstein tensor, and the last term can be written as $2P^\alpha_{\ \mu\nu}\pa_\alpha f_{Q}=2P^\alpha_{\ \mu\nu}f_{QQ}\partial_\alpha Q$.
Here, we use the GR-like form (see Eq. (145) in Ref. \cite{Capozziello2022})
\begin{equation} \label{eq:COND_EQ}
\lc{G}_{\mu\nu}=\frac{2}{e}\nabla_\alpha(2eP^\alpha{}_{\mu\nu})+\frac{1}{e}q_{\mu\nu}-\frac{1}{2} g_{\mu\nu}Q.
\end{equation}

If we derive the action with respect to the affine connection (i.e., $\delta_\Gamma S_Q=0$), we obtain the \emph{connection field equations} $\mathcal{C_\alpha}=0$, where \cite{DAmbrosio2022}
\begin{align} 
\mathcal{C_\alpha}&=\nabla_\mu\nabla_\nu\left[P^{\mu\nu}{}_\alpha e f_{Q}(Q)\right].\label{eq:CE}
\end{align}
The above equation can be found by introducing the Lagrange multipliers subjected to the constraints of vanishing torsion and curvature. Then, the hypermomentum can be defined as follows  \cite{Beltran2018}
\begin{equation} 
\mathfrak{H}^\lambda{}_{\mu\nu}=-\frac{1}{2}\frac{\partial\mathcal{L}_m}{\partial \Gamma^\alpha{}_{\mu\nu}}.
\end{equation} 
Requiring the hypermomentum conservation (i.e., $\nabla_\mu \nabla_\nu \mathfrak{H}_\alpha{}^{\mu\nu}=0$) and using the symmetry properties of the aforementioned Lagrange multipliers, we come to Eq. \eqref{eq:CE} (see Sec. IV--A in Ref. \cite{Beltran2018}, for more details). 

It is worth noticing that the conservation laws of the energy-momentum tensor with respect to the GR covariant divergence (i.e., $\lc{\nabla}^\mu \Theta_{\mu\nu}=0$) implies
\begin{align} \label{eq:cl_f(Q)}
\lc{\nabla}^\mu \mathcal{M}_{\mu\nu}&=\partial^\mu f_Q \left(\lc{G}_{\mu\nu}+\frac{1}{2}g_{\mu\nu}Q+2\lc{\nabla}^\lambda P_{\mu\lambda\nu}\right)\notag\\
&+2P_{\mu\lambda\nu}\lc{\nabla}^\lambda (\partial^\mu f_Q)=0,
\end{align} 
which is not identically satisfied, but it represents an additional constraint to be considered. Of course, Eq. \eqref{eq:cl_f(Q)}  holds in STEGR, as soon as  $f_Q=1$. 
 
\section{Improving the theory with a boundary term} 
\label{sec:f(Q,B)_gravity}
We propose an extension of $f(Q)$ gravity, by considering a generic smooth function of the non-metricity scalar and of the boundary term $B$, namely the $f(Q,B)$ gravity.
Let us start from the following action
\begin{equation} \label{eq:f(Q,B)_action}
S_{QB}=\int   \dd^4x \  e  \biggr{[}\frac{1}{2\chi}f(Q,B)+\mathcal{L}_m\biggr{]}. 
\end{equation}
Varying $S_{QB}$ with respect to the metric (i.e., $\delta_g S_{QB}=0$), we obtain the following metric field equations (see Appendix \ref{sec:f(Q,B)_gravity_VF}, for their derivations and details)
\begin{align} \label{eq:EFE1}
&\frac{2}{e}\nabla_\alpha(P^\alpha_{\ \mu\nu}ef_Q)+q_{\mu\nu}f_Q-\frac{1}{2}g_{\mu\nu}f+\frac{1}{2}g_{\mu\nu}f_B B\notag\\
&+(\partial_\lambda f_B)\biggr{[}\frac{1}{2}g_{\mu\nu}\tilde{Q}^\lambda-U^\lambda_{\mu\nu}\biggr{]}=\chi\Theta_{\mu\nu},
\end{align}
where (cf. Eq. \ref{eq:Ulun})
\begin{align}
(\partial_\lambda f_B)U^\lambda_{\ \mu\nu}&=\left[\Gamma^\lambda_{\ \mu\nu}-\frac{1}{2}\left(\delta^\lambda_\mu \Gamma^\rho_{\ \nu\rho}+\delta^\lambda_\nu \Gamma^\rho_{\ \mu\rho}\right)\right.\notag\\
&-\lc{\Gamma}^\lambda_{\ \mu\nu}\biggr{]}(\pa_\lambda f_B)+\frac{1}{2}\left(\pa_\nu f_B\lc{\Gamma}^\rho_{\ \mu\rho}+\pa_\mu f_B\lc{\Gamma}^\rho_{\ \nu\rho}\right)\notag\\
&-\lc{\nabla}_\mu\lc{\nabla}_\nu f_B +g_{\mu\nu}\lc{\Box}f_B.
\end{align}
Using Eq. \eqref{eq:COND_EQ}, the field equations \eqref{eq:EFE1} become
\begin{align} \label{eq:EFE2}
&\lc{G}_{\mu\nu} f_Q-\frac{1}{2}g_{\mu\nu}\biggr{(}f-f_QQ-f_B B\biggr{)}+(\partial_\lambda f_Q)2P^\lambda_{\ \mu\nu}\notag\\
&-(\partial_\lambda f_B)\biggr{[}-\frac{1}{2}g_{\mu\nu}\tilde{Q}^\lambda+U^\lambda_{\mu\nu}\biggr{]}=\chi\Theta_{\mu\nu}.
\end{align}
The above equation can be also written as 
\begin{align} \label{eq:EFE2_bis}
&\lc{G}_{\mu\nu} f_Q-\frac{1}{2}g_{\mu\nu}\biggr{(}f-f_QQ-f_B B\biggr{)}+\partial_\lambda(f_Q+f_B)2P^\lambda_{\ \mu\nu}\notag\\
&-(\partial_\lambda f_B)\biggr{[}2P^\lambda_{\ \mu\nu}-\frac{1}{2}g_{\mu\nu}\tilde{Q}^\lambda+U^\lambda_{\mu\nu}\biggr{]}=\chi\Theta_{\mu\nu}.
\end{align}
It is possible to prove that (see Appendix \ref{sec:f(Q,B)_gravity_VF})
\begin{equation} \label{eq:boundary_term_mean}
2P^\lambda_{\ \mu\nu}-\frac{1}{2}g_{\mu\nu}\tilde{Q}^\lambda+U^\lambda_{\mu\nu}=g_{\mu\nu}\lc{\Box}f_B-\lc{\nabla}_\mu\lc{\nabla}_\nu f_B.
\end{equation}
Therefore, Eq. \eqref{eq:EFE2} becomes
\begin{subequations}  \label{eq:EFE-final}
\begin{align} 
\mathcal{H}_{\mu\nu}&=\chi\Theta_{\mu\nu},\label{eq:EFE3}\\
\mathcal{H}_{\mu\nu}&=\lc{G}_{\mu\nu} f_Q-\frac{1}{2}g_{\mu\nu}\biggr{(}f-f_QQ-f_B B\biggr{)}\notag\\
&+\partial_\lambda(f_Q+f_B)2P^\lambda_{\ \mu\nu}-g_{\mu\nu}\lc{\Box}f_B+\lc{\nabla}_\mu\lc{\nabla}_\nu f_B.\label{eq:EFE3b}
\end{align}
\end{subequations}
It is important to note that the addition of a boundary term $B$ fulfills an important role, because it allows the $f(Q)$ gravity to ascend from second to fourth order field equations (cf. Eq. \eqref{eq:boundary_term_mean}). The reader can find a discussion on $f(T,B)$ gravity in Refs. \cite{Bahamonde2015,Sebastian,Capriolo}.

Adopting the same strategy employed in Sec. \ref{sec:f(Q)_gravity} to derive Eq. \eqref{eq:CE}, we finally obtain the connection field equation in $f(Q,B)$ gravity, namely
\begin{equation} \label{eq:ECE}
\nabla_\mu\nabla_\nu\left[P^{\mu\nu}{}_\alpha e(f_Q+f_B)\right]=0.
\end{equation}

\subsection{Consistency check}
\label{sec:consistency}
This section is dedicated to check the consistency of the ensued results, represented by Eqs. \eqref{eq:EFE-final} and \eqref{eq:ECE}. 

\begin{enumerate}
\item For $B=0$, we have $f_B=0$ and the field equations \eqref{eq:EFE-final} reduce to  $f(Q)$ gravity  \eqref{eq:FE}, as well as the connection equations  \eqref{eq:ECE} reduce to Eq. \eqref{eq:CE}.
\item For $R=Q-B$, we have $f(\lc{R})=f(Q-B)$ and $F(\lc{R})=f_R(\lc{R})=f_Q=-f_B$. From this preliminary analysis, we immediately note that Eq. \eqref{eq:EFE-final} is equivalent to the $f(R)$ gravity, namely \cite{Faraoni2010}
\begin{align}
&\lc{G}_{\mu\nu} F-\frac{1}{2}g_{\mu\nu}\biggr{(}f-F\lc{R}\biggr{)}\notag\\
&+g_{\mu\nu}\lc{\Box}F-\lc{\nabla}_\mu\lc{\nabla}_\nu F=\chi\Theta_{\mu\nu}.
\end{align}
Clearly, Eq. \eqref{eq:ECE} is trivial in this framework.  
\item Finally, we require that the stress-energy tensor is conserved under the action of the GR covariant divergence, namely $\lc{\nabla}^\mu \Theta_{\mu\nu}=0$, which implies 
\begin{align} \label{eq:cl_f(Q,B)}
\lc{\nabla}^\mu \mathcal{H}_{\mu\nu}&=\pa^\mu(f_Q+f_B)\biggr{[}\lc{R}_{\mu\nu}+\frac{1}{2}g_{\mu\nu}B+2\lc{\nabla}^\lambda P_{\lambda\mu\nu}\biggr{]}\notag\\
&+\lc{\nabla}^\lambda [\partial^\mu(f_Q+f_B)]2P_{\lambda\mu\nu}=0,
\end{align}
being not identically satisfied. This imposes thus a further constraint, as it already occurs in the $f(Q)$ gravity (see Eq. \eqref{eq:cl_f(Q)} and discussion below). However, it is important to note that Eq. \eqref{eq:cl_f(Q,B)} holds in the $f(Q-B)$ gravity, because $f_Q=-f_B$.
\end{enumerate}

\subsection{ The extension of Geometric Trinity of Gravity }
\label{sec:EGTG}
As already illustrated above, Geometric Trinity of Gravity gives three dynamically equivalent formulations of GR. They are based on Lagrangians containing $R,T,Q$, representing  the Ricci curvature, the torsion, and the non-metricity scalars, respectively. It is possible to demonstrate that these Lagrangians are equivalent up to a boundary term, which is different in GR, TEGR, and STEGR. This equivalence practically means that TEGR and STEGR have the same field equations of GR.

\begin{figure}
\includegraphics[scale=0.2]{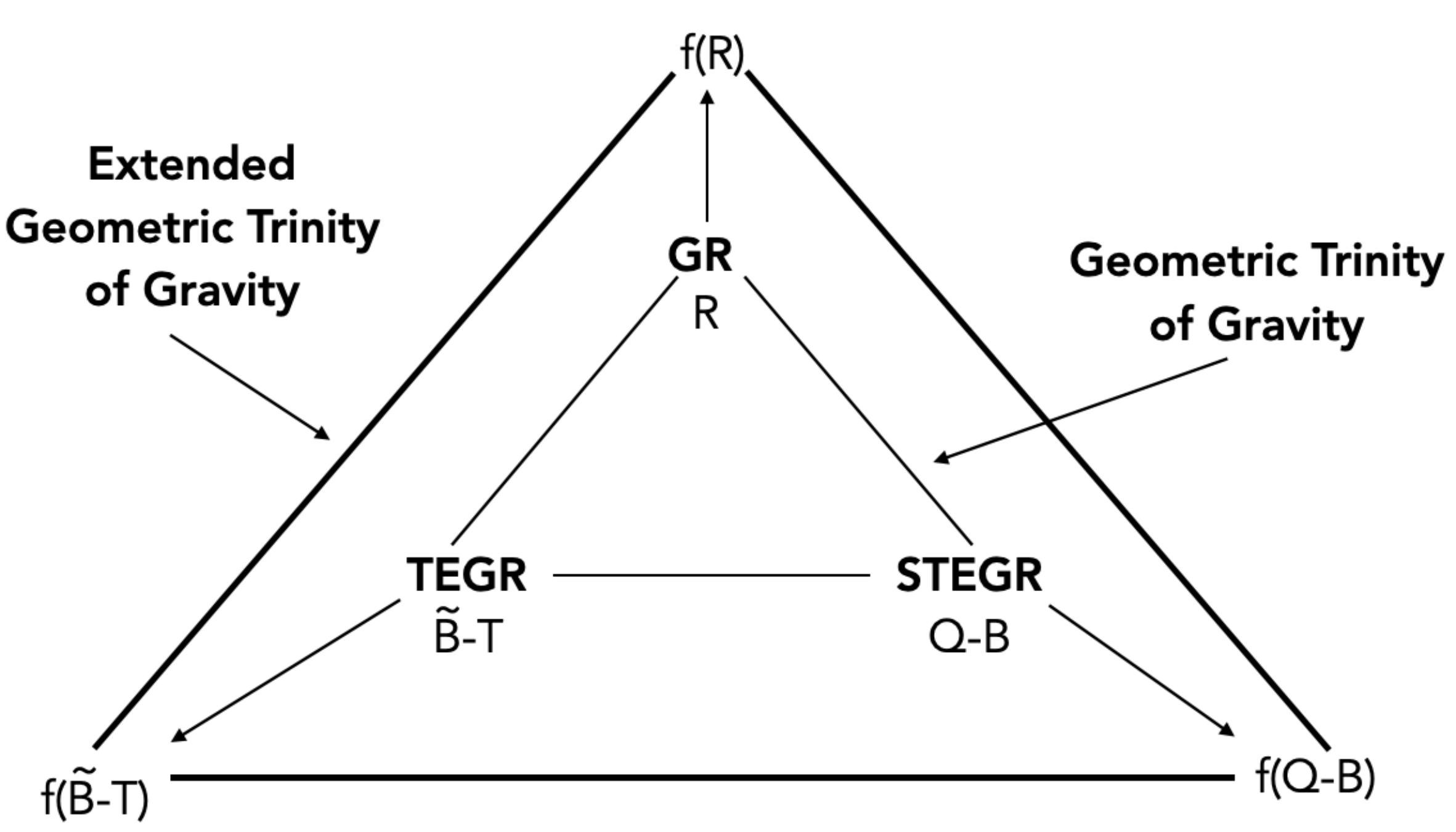}
\caption{Summary scheme of the Geometric Trinity of Gravity and Extended Geometric Trinity of Gravity.}
\label{fig:Fig1}
\end{figure}

However, the main issue arises, when we pass to the related extended theories, represented by $f(R),f(T),f(Q)$ gravities, which are not dynamically equivalent, because $f(R)$ gravity, in metric formalism, is a fourth-order theory, whereas  $f(T)$ and $f(Q)$ are second-order theories. Nevertheless, it is possible to restore the equivalence among extended theories via the addition of an appropriate boundary term. Indeed, in the general frameworks $f(T,\tilde{B})$ and $f(Q,B)$ (where usually $\tilde{B}\neq B$), we have that $f(\tilde{B}-T),f(Q-B)$ are dynamically equivalent to $f(R)$ gravity. These three theories constitute what we may dub \emph{Extended Geometric Trinity of Gravity} (see Fig. \ref{fig:Fig1}). While $R,\tilde{B}-T,Q-B$ is a geometric trinity of gravity of second-order, $f(R), f(\tilde{B}-T),f(Q-B)$ configures to be as a geometric trinity of gravity of fourth-order. In summary, adding  boundary terms means to improve the number of degrees of freedom, because they act as effective scalar fields. Clearly, this procedure can be extended to higher-order metric-affine theories formulated in metric, teleparallel, and symmetric teleparallel formalisms.

\subsection{The role of boundary terms in GR and $f(Q,B)$}
It is important to note that in GR there exists the \emph{Gibbons-Hawking-York (GHY) boundary term} $S_{\rm GHY}$ \cite{Gibbons1977}, which becomes fundamental when the underlying spacetime manifold $\mathcal{M}$ is endowed with a boundary $\partial\mathcal{M}$. In this case, in order to recover the Einstein field equations, we must add it to the Einstein-Hilbert action. The hypersurface $\partial\mathcal{M}$ can be described by parametric equations $x^\alpha=x^\alpha(y^i)$ with $i=1,2,3$ and $y^i$ are the intrinsic coordinates on $\partial\mathcal{M}$. Then, we define the unit normal $n^\alpha$ to the hypersurface $\partial\mathcal{M}$ as follows\footnote{We consider the metric signature $(-,+,+,+)$.}
\begin{equation}  
n^\alpha n_\alpha=\epsilon =\begin{cases}
-1 & \mbox{if $\partial\mathcal{M}$ is timelike},\\
+1 & \mbox{if $\partial\mathcal{M}$ is spacelike}.
\end{cases}
\end{equation}  
It is useful also to introduce the tangent vectors $t^\alpha_a$ to $\partial\mathcal{M}$ (being orthogonal to $n^\alpha$, i.e., $n_\alpha t^\alpha_a=0$) as follows\footnote{The index $\alpha$ labels the components, whereas $a$ counts the independent tangent vectors, namely $t^\alpha_1,t^\alpha_2,t^\alpha_3$.}
\begin{equation}
t^\alpha_a=\left(\frac{\partial x^\alpha}{\partial y^a}\right)_{\partial\mathcal{M}},\quad a=1,2,3;\quad \alpha=0,1,2,3.
\end{equation}
The previous concepts permit to define the induced metric $h_{ab}$ and the trace of the extrinsic (or Gaussian) curvature $K$ on the hypersurface $\partial\mathcal{M}$, which read explicitly as \cite{Romano2019}
\begin{align}
h_{ab}&=g_{\alpha\beta}t^\alpha_a t^\beta_b,\\
K&=\lc{\nabla}_\alpha n^\alpha.
\end{align}
We have now all the ingredients to define the GHY term
\begin{equation}
S_{\rm GHY}=\frac{1}{8\pi}\int_{\partial \mathcal{M}} \dd^3 y \epsilon \sqrt{|h|} K.
\end{equation}
The above quantity has the objective to cancel out the following GR divergence piece \cite{Romano2019} 
\begin{equation}
\int_{\mathcal{M}}g^{\alpha\beta} \delta R_{\alpha\beta} e\dd^4 x=\int_{\mathcal{M}}\nabla_\mu \delta W^\mu e \dd^4 x,
\end{equation}
where $\delta W^\mu=g^{\alpha\beta}\delta\Gamma^\mu_{\alpha\beta}-g^{\alpha\mu}\delta\Gamma^\beta_{\alpha\beta}$, which vanishes only in the case of a manifold without boundary. The GHY boundary term is also largely employed for path integral to quantum gravity and for calculating the black hole entropy via the Euclidean semi-classical approach \cite{Dyer2009}. To summarise we can regard the GHY boundary term having a \emph{conservative action} for assuring the validity of GR field equations on a manifold with a boundary.

Instead, the boundary term $B$ in $f(Q,B)$ (as well as $\tilde{B}$ in $f(T,\tilde{B})$) permits to upgrade the theory from a second-order to fourth-order in the field equations. In particular choosing $f(Q,B)=f(Q-B)$ (or $f(T,\tilde{B})=f(\tilde{B}-T)$) we obtain a special theory, which is equivalent to $f(R)$. In particular for $f(Q-B)$ holds a series of interesting properties explained in Sec. \ref{sec:consistency}. In addition, the boundary term $B$ (cf. Eq. \eqref{eq:boundary_term}), as well as $\tilde{B}$, is a second-order invariant. In this case the boundary term fulfills an \emph{upgrade action} on the underlying theory of gravity.

Although $S_{\rm GHY}$ and $B$ are employed for achieving different goals, there is a contact point, when we consider the $f(Q,B)$ ($f(T,\tilde{B})$) framework settled on a manifold $\mathcal{M}$ endowed with a boundary $\partial \mathcal{M}$. In this case, a GHY-like boundary term, depending on $B$ ($\tilde{B}$), should be added in order to recover the appropriate field equations.

\section{Discussion and conclusions}
\label{sec:end}
In this paper, we have discussed $f(Q,B)$ gravity, which is an extension  of Symmetric Teleparallel Gravity endowed with some interesting properties. Thanks to the introduction of an appropriate boundary term $B$, it is possible to lift up the $f(Q)$ gravity from second-order to fourth-order in the field equations. In particular, the model $f(Q,B)=f(Q-B)$ is dynamically equivalent to $f(R)$ gravity. Furthermore, the theory is consistent with $f(Q)$ gravity in the limit of $B\to0$. Furthermore, this theory identically satisfy the conservation laws with respect to the GR divergence. We can say that $f(R),f(\tilde{B}-T),f(Q-B)$ (with $\tilde{B}\neq B$) give rise to an extension of trinity gravity, which we have dubbed the {\it Extended Geometric Trinity of Gravity}, being three equivalent theories based on fourth-order field equations. 

We have also discussed the difference existing between the GHY boundary term in GR and the boundary term $B$ in $f(Q,B)$. The former has a conservative action, because it must reproduce the Einstein field equations on a manifold endowed with a boundary, whereas the latter plays a role of theory's improvement, since it lifts the field equations from the second-order to the fourth-order.

It is worth noticing that these results provide a sort of route to extend equivalent representations of gravity to any higher-order dynamics (see also Ref. \cite{Capriolo2}, for details), which can be extremely interesting in cosmological and astrophysical applications. As mentioned above, being $f(R)$ gravity of fourth-order in metric formalism, it seems that it cannot be directly compared with $f(T)$ and $f(Q)$, which are of second-order. If one considers cosmology, for example, it seems that dynamics coming from Starobinsky gravity \cite{Starobinsky}, i.e. $f(R)=R+\alpha R^2$, cannot be theoretically compared with $f(T)=T+\alpha_1 T^2$ or $f(Q)=Q+\alpha_2 Q^2$ \cite{Shokri} due to the different order of related field equations. The introduction of boundary terms restores the same differential order into the equations and then information coming from different representations of gravity appears on the same ground. Similar considerations can be developed also for black hole solutions or, in more general, for any self-gravitating compact object.

In a forthcoming paper, we will study how boundary terms impact phenomenology. In particular, it is relatively straightforward to show that the matching of cosmological models against the observations may result improved thanks to the presence of boundary terms.

\section*{Acknowledgements}
This paper is based upon work from COST Action CA21136 Addressing observational tensions in cosmology with systematics and fundamental physics (CosmoVerse) supported by COST (European Cooperation in Science and Technology). Authors acknowledge the Istituto Nazionale di
Fisica Nucleare (iniziative specifiche QGSKY, TEONGRAV and MOONLIGHT2).
S.C. and V.D.F. thank the  Gruppo Nazionale di Fisica Matematica of Istituto Nazionale di Alta Matematica for the support.  We are  grateful to  Francesco Bajardi and Rocco D'Agostino for useful discussions and suggestions.

\appendix
\section{Variational principle for $f(Q)$ gravity and field equations}
\label{sec:f(Q)_gravity_VF}
Starting from Eq. \eqref{eq:f(Q)_action} and considering the variation of the action with respect to the metric, we have
\begin{align} \label{eq:starting_eq}
\delta_g(ef)&=\biggr{[}\del{e} f+ e f_Q \del{Q}\biggr{]}\delta_g g^{\mu\nu},
\end{align}
where\footnote{The variation $\pg{e}=-\frac{1}{2}g_{\mu\nu}e$ can be obtained via the \emph{Jacobi formula}, expressing the derivative of the determinant of a matrix $A$ in terms of the adjugate of $A$ and the derivative of $A$ \cite{Magnus1999}.}
\begin{subequations}
\begin{align}
f\del{e}&=-\frac{1}{2}g_{\mu\nu}ef \del g^{\mu\nu},\\
ef_Q \del{Q}&=ef_Q\left[\pg{Q}\del g^{\mu\nu}+\frac{\partial Q}{\partial Q_\alpha^{\ \mu\nu}}\del Q_\alpha^{\ \mu\nu}\right]\notag\\
&-\pg{e}f_Q Q\del g^{\mu\nu}-e\pg{f_Q}Q\del g^{\mu\nu}\notag\\
&+ef_Q(2P^\alpha_{\ \mu\nu})(-\nabla_\alpha\del g^{\mu\nu})\notag\\
&=\Biggr{[}\frac{1}{2}g_{\mu\nu}f_QQ+f_Q\frac{1}{e}\frac{\pa (eQ)}{\pa g^{\mu\nu}}\notag\\
&\qquad+\frac{1}{e}\nabla(2ef_QP^\alpha_{\ \mu\nu})\Biggr{]}e\del g^{\mu\nu}\notag\\
&=\Biggr{[}\frac{1}{e}q_{\mu\nu}f_Q+\frac{1}{e}\nabla(2ef_QP^\alpha_{\ \mu\nu})\Biggr{]}e\del g^{\mu\nu}.
\end{align}
\end{subequations}
In the above implications we have used the definition of $P^\alpha_{\ \mu\nu}$ (cf. Eq. \eqref{eq:P})\footnote{In Eq. \eqref{eq:equ2}, we have used the following relation
\begin{align}
Q_\alpha^{\ \mu\nu}&=g^{\mu\beta}g^{\nu\gamma}Q_{\alpha\beta\gamma}=g^{\mu\beta}g^{\nu\gamma}(\pa_\alpha g_{\beta\gamma}-\Gamma^\lambda_{\ \alpha\beta}g_{\lambda\gamma}-\Gamma^\lambda_{\ \alpha\gamma}g_{\lambda\beta})\notag\\
&=\pa_\alpha(g^{\mu\beta}g^{\nu\gamma} g_{\beta\gamma})- g^{\nu\gamma} g_{\beta\gamma}\pa_\alpha g^{\mu\beta}- g^{\mu\beta} g_{\beta\gamma}\pa_\alpha g^{\nu\gamma}\notag\\
&-g^{\mu\beta}g^{\nu\gamma}g_{\lambda\gamma}\Gamma^\lambda_{\ \alpha\beta}-g^{\mu\beta}g^{\nu\gamma}g_{\lambda\beta}\Gamma^\lambda_{\ \alpha\gamma}\notag\\
&=-(\pa_\alpha g^{\mu\nu}+\Gamma^\nu_{\alpha\beta}g^{\mu\beta}+\Gamma^\mu_{\alpha\beta}g^{\nu\beta})=-\nabla_\alpha g^{\mu\nu}. \label{eq:equ2}
\end{align}}, and $q_{\mu\nu}$ (see Eq. \eqref{eq:qmunu}). For an alternative calculation of $\del Q$, we suggest the reader to see Appendices A and B in Ref. \cite{Xu2019}.

Inserting the above findings in Eq. \eqref{eq:starting_eq}, we obtain 
\begin{align}
\frac{2}{e}\nabla_\alpha(eP^\alpha_{\ \mu\nu}f_Q)+\frac{1}{e}q_{\mu\nu}f_Q-\frac{1}{2}g_{\mu\nu}f=\chi \Theta_{\mu\nu}.
\end{align}

\section{Variational principle for $f(Q,B)$ gravity and field equations}
\label{sec:f(Q,B)_gravity_VF}
Starting from the modified action \eqref{eq:f(Q,B)_action} and varying it with respect to the metric, we obtain
\begin{subequations} \label{eq:start}
\begin{align} 
\delta_g(e\mathcal{L}_m)&=-e\chi\Theta_{\mu\nu}\del g^{\mu\nu}, \label{eq:start_a}\\
\delta_g(ef)&=\del e f+e f_Q\del{Q}+ ef_B \del{B}, \label{eq:start_b}
\end{align} 
\end{subequations} 
where
\begin{subequations} \label{eq:second}
\begin{align}
&\del e f+e f_Q\del{Q}=e\mathcal{M}_{\mu\nu},\\
&ef_B \del{B}=ef_B\del\left[\frac{1}{e}\partial_\lambda (e\tilde{Q}^\lambda)\right]\notag\\
&=\frac{1}{2}g_{\mu\nu}e f_B B\del g^{\mu\nu}+(\pa_\lambda f_B)\left(\frac{1}{2}g_{\mu\nu}\tilde{Q}^\lambda\right)e\del g^{\mu\nu}\notag\\
&\quad-(\pa_\lambda f_B)(e\del \tilde{Q}^\lambda)\notag\\
&=\biggr{[}\frac{g_{\mu\nu}}{2}f_B B +(\partial_\lambda f_B)\biggr{(}\frac{g_{\mu\nu}}{2}\tilde{Q}^\lambda-U^\lambda_{\mu\nu}\biggr{)}\biggr{]}e\del g^{\mu\nu},
\end{align}
\end{subequations} 
with $e(\pa_\lambda f_B)\del \tilde{Q}^\lambda=eU^\lambda_{\ \mu\nu} \del g^{\mu\nu}$. Let us explicitly calculate this last term, where we obtain
\begin{align} \label{eq:first_term}
e(\pa_\lambda f_B)\del \tilde{Q}^\lambda&=\del{}\biggr{[}g^{\lambda\alpha}g^{\sigma\rho}(\partial_\alpha g_{\sigma\rho}-\partial_\rho g_{\sigma\alpha})\notag\\
&-g^{\lambda\alpha}\Gamma^\rho_{\ \alpha\rho}+g^{\sigma\rho}\Gamma^\lambda_{\ \sigma\rho}\biggr{]}e(\partial_\lambda f_B)\notag\\
&=\underbrace{e(\partial_\lambda f_B)\left[\del{(g^{\lambda\alpha}g^{\sigma\rho})}\right](\partial_\alpha g_{\sigma\rho}-\partial_\rho g_{\sigma\alpha})}_{(A_1)}\notag\\
&+\underbrace{e(\partial_\lambda f_B)g^{\lambda\alpha}g^{\sigma\rho}\left[\del{}(\partial_\alpha g_{\sigma\rho}-\partial_\rho g_{\sigma\alpha})\right]}_{(A_2)}\notag\\
&+\underbrace{e(\partial_\lambda f_B)\del{}(g^{\sigma\rho}\Gamma^\lambda_{\ \sigma\rho}-g^{\lambda\alpha}\Gamma^\rho_{\ \alpha\rho})}_{(A_3)}.
\end{align}
Calculating all the above terms, we eventually have\footnote{We remind that the Levi-Civita affine connection $\lc{\Gamma}^\lambda_{\ \mu\nu}$ is
\begin{equation}
\lc{\Gamma}^\lambda_{\ \mu\nu}=\frac{1}{2}g^{\lambda\alpha}\left(\partial_\mu g_{\alpha\nu}+\partial_\nu g_{\alpha\mu}-\partial_\alpha g_{\mu \nu}\right).
\end{equation}}
\begin{align}
A_1&=\left\{-\lc{\Gamma}^\lambda_{\ \mu\nu}+\frac{1}{2}\partial_\alpha g_{\mu\nu} g^{\lambda\alpha}+\frac{1}{2}g^{\sigma\rho}(\delta^\lambda_\mu \partial_\nu g_{\sigma\rho}+\delta^\lambda_\nu \partial_\mu g_{\sigma\rho})\right.\notag\\
&\left.-\frac{1}{2}g^{\sigma\rho}(\delta^\lambda_\mu \partial_\rho g_{\sigma\nu}+\delta^\lambda_\nu \partial_\rho g_{\sigma\mu})\right\}e(\pa_\lambda f_B)\del g^\mu\nu,\\
A_2&=\left\{\Biggr{[}-2g^{\lambda\alpha}\partial_\alpha g_{\mu\nu}+\frac{1}{2}g^{\sigma\rho}\left(\delta^\lambda_\nu \partial_\rho g_{\mu\sigma}+\delta^\lambda_\mu\partial_\rho g_{\nu\sigma}\right)\right.\notag\\
&+\frac{1}{2}g^{\lambda\alpha}(\partial_\mu g_{\nu\alpha}+\partial_\nu g_{\mu\alpha})+\pa_\alpha(g^{\lambda\alpha}g_{\mu\nu})\notag\\
& +\lc{\Gamma}^\rho_{\ \alpha\rho}g^{\lambda\alpha}g_{\mu\nu}\Biggr{]}(\pa_\lambda f_B)+g^{\alpha\beta}g_{\mu\nu}\pa_\alpha\pa_\beta f_B -\pa_\mu \pa_\nu f_B\notag\\
&\left.-\frac{1}{2}\left(\pa_\nu f_B \lc{\Gamma}^\rho_{\ \mu\rho}+\pa_\mu f_B \lc{\Gamma}^\rho_{\ \nu\rho}\right)\right\}e\del g^{\mu\nu},\\
A_3&=\left\{\Gamma^\lambda_{\ \mu\nu}-\frac{1}{2}\left(\delta^\lambda_\mu \Gamma^\rho_{\ \nu\rho}+\delta^\lambda_\nu \Gamma^\rho_{\ \mu\rho}\right)\right\}e(\pa_\lambda f_B)\del g^{\mu\nu},
\end{align}
where in $A_1$ and $A_3$ we have employed 
\begin{align}
\pg{g^{\rho\sigma}}=\frac{1}{2}\left(\delta^{\mu\rho}\delta^{\nu\sigma}+\delta^{\nu\rho}\delta^{\mu\sigma}\right);
\end{align}
instead in $A_2$ we have exchanged $\partial_\alpha$ and $\delta_g$ and used
\begin{subequations}
\begin{align}
\del g_{\rho\sigma}&=-\frac{1}{2}\left(g_{\rho\mu}g_{\sigma\nu}+g_{\sigma\mu}g_{\rho\nu}\right)\del g^{\mu\nu},\\
\del e&=e \lc{\Gamma}^\rho_{\ \alpha\rho}.
\end{align}
\end{subequations}

Therefore, from Eqs. \eqref{eq:first_term} we have (without $e\del g^{\mu\nu}$)
\begin{align} \label{eq:Ulun}
(\partial_\lambda f_B)U^\lambda_{\ \mu\nu}&=A_1+A_2+A_3\notag\\
&=\left[\Gamma^\lambda_{\ \mu\nu}-\frac{1}{2}\left(\delta^\lambda_\mu \Gamma^\rho_{\ \nu\rho}+\delta^\lambda_\nu \Gamma^\rho_{\ \mu\rho}\right)\right.\notag\\
&-\lc{\Gamma}^\lambda_{\ \mu\nu}\biggr{]}(\pa_\lambda f_B)+\frac{1}{2}\left(\pa_\nu f_B\lc{\Gamma}^\rho_{\ \mu\rho}+\pa_\mu f_B\lc{\Gamma}^\rho_{\ \nu\rho}\right)\notag\\
&-\lc{\nabla}_\mu\lc{\nabla}_\nu f_B +g_{\mu\nu}\lc{\Box}f_B,
\end{align}
where we have used the following identities
\begin{subequations}
\begin{align}
&\partial_\alpha g_{\beta\gamma}=\lc{\Gamma}^\rho_{\ \alpha\beta}g_{\rho\gamma}+\lc{\Gamma}^\rho_{\ \alpha\gamma}g_{\rho\beta},\\
&\partial_\alpha g^{\beta\gamma}=-g^{\beta\gamma}g^{\gamma\sigma}\partial_\alpha g_{\rho\sigma},\\
&\frac{1}{2}g^{\sigma\rho}(\partial_\lambda f_B)(\delta^\lambda_\mu\pa_\nu g_{\sigma\rho}+\delta^\lambda_\nu\pa_\mu g_{\sigma\rho})\notag\\
&=\frac{1}{2}\left[\pa_\mu f_B\lc{\Gamma}^\rho_{\ \nu\rho}+\pa_\nu f_B\lc{\Gamma}^\rho_{\ \mu\rho}\right].
\end{align}
\end{subequations}

Starting from Eqs. \eqref{eq:start} and considering Eqs. \eqref{eq:second} and \eqref{eq:Ulun}, we obtain the following metric field equations \eqref{eq:EFE1}. Now, we calculate the following terms 
\begin{align} \label{eq:P+Q}
2P^\lambda_{\ \mu\nu}-\frac{1}{2}g_{\mu\nu}\tilde{Q}^\lambda&=-\frac{1}{2}Q^\lambda_{\ \mu\nu}+\frac{1}{2}\left(Q_\mu{}^\lambda_{\ \nu}-Q_\nu{}^\lambda_{\ \mu}\right)\notag\\
&-\frac{1}{4}\left(\delta^\lambda_\mu Q_\nu+\delta^\lambda_\nu Q_\mu\right)\notag\\
&=\lc{\Gamma}^\lambda_{\ \mu\nu}-\frac{1}{4}\left(\delta^\lambda_\mu\partial_\nu g_{\sigma\rho}+\delta^\lambda_\nu\partial_\mu g_{\sigma\rho}\right)g^{\sigma\rho}\notag\\
&-\Gamma^\lambda_{\ \mu\nu}+\frac{1}{2}\left(\delta^\lambda_\mu \Gamma^\alpha_{\ \nu\alpha}+\delta^\lambda_\nu \Gamma^\alpha_{\ \mu\alpha}\right)\notag\\
&=\lc{\Gamma}^\lambda_{\ \mu\nu}-\frac{1}{2}\left(\pa_\nu f_B\lc{\Gamma}^\rho_{\ \mu\rho}+\pa_\mu f_B\lc{\Gamma}^\rho_{\ \nu\rho}\right)\notag\\
&-\Gamma^\lambda_{\ \mu\nu}+\frac{1}{2}\left(\delta^\lambda_\mu \Gamma^\alpha_{\ \nu\alpha}+\delta^\lambda_\nu \Gamma^\alpha_{\ \mu\alpha}\right).
\end{align}

After simple algebra, we obtain Eq. \eqref{eq:EFE2_bis}. Gathering Eqs. \eqref{eq:Ulun} and \eqref{eq:P+Q}, we prove Eq. \eqref{eq:boundary_term_mean}. Defined $\Psi_{\mu\nu}=(\partial_\lambda f_B)(2P^\lambda_{\ \mu\nu}-\frac{1}{2}g_{\mu\nu}\tilde{Q}^\lambda+U^\lambda_{\ \mu\nu})$, we have
\begin{align} \label{eq:Psi_1}
\Psi_{\mu\nu}&=g_{\mu\nu}\lc{\Box}f_B-\lc{\nabla}_\mu\lc{\nabla}_\nu f_B.
\end{align}
The field equations of $f(Q,B)$ gravity are (cf. Eq. \eqref{eq:EFE-final})
\begin{align} 
&\lc{G}_{\mu\nu} f_Q-\frac{1}{2}g_{\mu\nu}\biggr{(}f-f_QQ-f_B B\biggr{)}+\partial_\lambda(f_Q\notag\\
&+f_B)2P^\lambda_{\ \mu\nu}-g_{\mu\nu}\lc{\Box}f_B+\lc{\nabla}_\mu\lc{\nabla}_\nu f_B=\chi\Theta_{\mu\nu}.
\end{align}


\begin{thebibliography}{45}%
\makeatletter
\providecommand \@ifxundefined [1]{%
 \@ifx{#1\undefined}
}%
\providecommand \@ifnum [1]{%
 \ifnum #1\expandafter \@firstoftwo
 \else \expandafter \@secondoftwo
 \fi
}%
\providecommand \@ifx [1]{%
 \ifx #1\expandafter \@firstoftwo
 \else \expandafter \@secondoftwo
 \fi
}%
\providecommand \natexlab [1]{#1}%
\providecommand \enquote  [1]{``#1''}%
\providecommand \bibnamefont  [1]{#1}%
\providecommand \bibfnamefont [1]{#1}%
\providecommand \citenamefont [1]{#1}%
\providecommand \href@noop [0]{\@secondoftwo}%
\providecommand \href [0]{\begingroup \@sanitize@url \@href}%
\providecommand \@href[1]{\@@startlink{#1}\@@href}%
\providecommand \@@href[1]{\endgroup#1\@@endlink}%
\providecommand \@sanitize@url [0]{\catcode `\\12\catcode `\$12\catcode
  `\&12\catcode `\#12\catcode `\^12\catcode `\_12\catcode `\%12\relax}%
\providecommand \@@startlink[1]{}%
\providecommand \@@endlink[0]{}%
\providecommand \url  [0]{\begingroup\@sanitize@url \@url }%
\providecommand \@url [1]{\endgroup\@href {#1}{\urlprefix }}%
\providecommand \urlprefix  [0]{URL }%
\providecommand \Eprint [0]{\href }%
\providecommand \doibase [0]{http://dx.doi.org/}%
\providecommand \selectlanguage [0]{\@gobble}%
\providecommand \bibinfo  [0]{\@secondoftwo}%
\providecommand \bibfield  [0]{\@secondoftwo}%
\providecommand \translation [1]{[#1]}%
\providecommand \BibitemOpen [0]{}%
\providecommand \bibitemStop [0]{}%
\providecommand \bibitemNoStop [0]{.\EOS\space}%
\providecommand \EOS [0]{\spacefactor3000\relax}%
\providecommand \BibitemShut  [1]{\csname bibitem#1\endcsname}%
\let\auto@bib@innerbib\@empty
\bibitem [{\citenamefont {Carroll}(2001)}]{Carroll2001}%
  \BibitemOpen
  \bibfield  {author} {\bibinfo {author} {\bibfnamefont {S.~M.}\ \bibnamefont
  {Carroll}},\ }\href {\doibase 10.12942/lrr-2001-1} {\bibfield  {journal}
  {\bibinfo  {journal} {Living Rev. Rel.}\ }\textbf {\bibinfo {volume} {4}},\
  \bibinfo {pages} {1} (\bibinfo {year} {2001})},\ \Eprint
  {http://arxiv.org/abs/astro-ph/0004075} {arXiv:astro-ph/0004075} \BibitemShut
  {NoStop}%
\bibitem [{\citenamefont {{Capozziello}}\ \emph {et~al.}(2019)\citenamefont
  {{Capozziello}}, \citenamefont {{D'Agostino}},\ and\ \citenamefont
  {{Luongo}}}]{Rocco2019}%
  \BibitemOpen
  \bibfield  {author} {\bibinfo {author} {\bibfnamefont {S.}~\bibnamefont
  {{Capozziello}}}, \bibinfo {author} {\bibfnamefont {R.}~\bibnamefont
  {{D'Agostino}}}, \ and\ \bibinfo {author} {\bibfnamefont {O.}~\bibnamefont
  {{Luongo}}},\ }\href {\doibase 10.1142/S0218271819300167} {\bibfield
  {journal} {\bibinfo  {journal} {International Journal of Modern Physics D}\
  }\textbf {\bibinfo {volume} {28}},\ \bibinfo {eid} {1930016} (\bibinfo {year}
  {2019})},\ \Eprint {http://arxiv.org/abs/1904.01427} {arXiv:1904.01427
  [gr-qc]} \BibitemShut {NoStop}%
\bibitem [{\citenamefont {{Planck Collaboration}}\ \emph
  {et~al.}(2020)\citenamefont {{Planck Collaboration}}, \citenamefont
  {{Aghanim}} \emph {et~al.}}]{Planckcollaboration2020}%
  \BibitemOpen
  \bibfield  {author} {\bibinfo {author} {\bibnamefont {{Planck
  Collaboration}}}, \bibinfo {author} {\bibfnamefont {N.}~\bibnamefont
  {{Aghanim}}},  \emph {et~al.},\ }\href {\doibase 10.1051/0004-6361/201833910}
  {\bibfield  {journal} {\bibinfo  {journal} {A\&A}\ }\textbf {\bibinfo
  {volume} {641}},\ \bibinfo {eid} {A6} (\bibinfo {year} {2020})},\ \Eprint
  {http://arxiv.org/abs/1807.06209} {arXiv:1807.06209 [astro-ph.CO]}
  \BibitemShut {NoStop}%
\bibitem [{\citenamefont {'t~Hooft}\ and\ \citenamefont
  {Veltman}(1974)}]{tHooft1974}%
  \BibitemOpen
  \bibfield  {author} {\bibinfo {author} {\bibfnamefont {G.}~\bibnamefont
  {'t~Hooft}}\ and\ \bibinfo {author} {\bibfnamefont {M.~J.~G.}\ \bibnamefont
  {Veltman}},\ }\href@noop {} {\bibfield  {journal} {\bibinfo  {journal} {Ann.
  Inst. H. Poincare Phys. Theor. A}\ }\textbf {\bibinfo {volume} {20}},\
  \bibinfo {pages} {69} (\bibinfo {year} {1974})}\BibitemShut {NoStop}%
\bibitem [{\citenamefont {Goroff}\ and\ \citenamefont
  {Sagnotti}(1985)}]{Goroff1985}%
  \BibitemOpen
  \bibfield  {author} {\bibinfo {author} {\bibfnamefont {M.~H.}\ \bibnamefont
  {Goroff}}\ and\ \bibinfo {author} {\bibfnamefont {A.}~\bibnamefont
  {Sagnotti}},\ }\href {\doibase 10.1016/0370-2693(85)91470-4} {\bibfield
  {journal} {\bibinfo  {journal} {Phys. Lett. B}\ }\textbf {\bibinfo {volume}
  {160}},\ \bibinfo {pages} {81} (\bibinfo {year} {1985})}\BibitemShut
  {NoStop}%
\bibitem [{\citenamefont {{'t Hooft}}\ and\ \citenamefont
  {{Veltman}}(1974)}]{Hooft1974}%
  \BibitemOpen
  \bibfield  {author} {\bibinfo {author} {\bibfnamefont {G.}~\bibnamefont {{'t
  Hooft}}}\ and\ \bibinfo {author} {\bibfnamefont {M.}~\bibnamefont
  {{Veltman}}},\ }\href@noop {} {\bibfield  {journal} {\bibinfo  {journal}
  {Annales de L'Institut Henri Poincare Section (A) Physique Theorique}\
  }\textbf {\bibinfo {volume} {20}},\ \bibinfo {pages} {69} (\bibinfo {year}
  {1974})}\BibitemShut {NoStop}%
\bibitem [{\citenamefont {{Gleiser}}(2005)}]{Gleiser2005}%
  \BibitemOpen
  \bibfield  {author} {\bibinfo {author} {\bibfnamefont {M.}~\bibnamefont
  {{Gleiser}}},\ }\href@noop {} {\bibfield  {journal} {\bibinfo  {journal}
  {Physics Today}\ }\textbf {\bibinfo {volume} {58}},\ \bibinfo {pages} {57}
  (\bibinfo {year} {2005})}\BibitemShut {NoStop}%
\bibitem [{\citenamefont {{Clifton}}(2006)}]{Clifton2006}%
  \BibitemOpen
  \bibfield  {author} {\bibinfo {author} {\bibfnamefont {T.}~\bibnamefont
  {{Clifton}}},\ }\emph {\bibinfo {title} {{Alternative Theories of
  Gravity}}},\ \href@noop {} {Ph.D. thesis},\ \bibinfo  {school} {-} (\bibinfo
  {year} {2006})\BibitemShut {NoStop}%
\bibitem [{\citenamefont {Capozziello}\ and\ \citenamefont
  {Faraoni}(2011)}]{Faraoni2010}%
  \BibitemOpen
  \bibfield  {author} {\bibinfo {author} {\bibfnamefont {S.}~\bibnamefont
  {Capozziello}}\ and\ \bibinfo {author} {\bibfnamefont {V.}~\bibnamefont
  {Faraoni}},\ }\href {\doibase 10.1007/978-94-007-0165-6} {\emph {\bibinfo
  {title} {{Beyond Einstein Gravity}: {A Survey of Gravitational Theories for
  Cosmology and Astrophysics}}}}\ (\bibinfo  {publisher} {Springer},\ \bibinfo
  {address} {Dordrecht},\ \bibinfo {year} {2011})\BibitemShut {NoStop}%
\bibitem [{\citenamefont {Nojiri}\ and\ \citenamefont
  {Odintsov}(2011)}]{Odintsov2011}%
  \BibitemOpen
  \bibfield  {author} {\bibinfo {author} {\bibfnamefont {S.}~\bibnamefont
  {Nojiri}}\ and\ \bibinfo {author} {\bibfnamefont {S.~D.}\ \bibnamefont
  {Odintsov}},\ }\href {\doibase 10.1016/j.physrep.2011.04.001} {\bibfield
  {journal} {\bibinfo  {journal} {Phys. Rept.}\ }\textbf {\bibinfo {volume}
  {505}},\ \bibinfo {pages} {59} (\bibinfo {year} {2011})},\ \Eprint
  {http://arxiv.org/abs/1011.0544} {arXiv:1011.0544 [gr-qc]} \BibitemShut
  {NoStop}%
\bibitem [{\citenamefont {{Capozziello}}\ and\ \citenamefont {{de
  Laurentis}}(2011)}]{Capozziello2011R}%
  \BibitemOpen
  \bibfield  {author} {\bibinfo {author} {\bibfnamefont {S.}~\bibnamefont
  {{Capozziello}}}\ and\ \bibinfo {author} {\bibfnamefont {M.}~\bibnamefont
  {{de Laurentis}}},\ }\href {\doibase 10.1016/j.physrep.2011.09.003}
  {\bibfield  {journal} {\bibinfo  {journal} {Phys. Rep.}\ }\textbf {\bibinfo
  {volume} {509}},\ \bibinfo {pages} {167} (\bibinfo {year} {2011})},\ \Eprint
  {http://arxiv.org/abs/1108.6266} {arXiv:1108.6266 [gr-qc]} \BibitemShut
  {NoStop}%
\bibitem [{\citenamefont {Nojiri}\ \emph {et~al.}(2017)\citenamefont {Nojiri},
  \citenamefont {Odintsov},\ and\ \citenamefont {Oikonomou}}]{Oikonomou}%
  \BibitemOpen
  \bibfield  {author} {\bibinfo {author} {\bibfnamefont {S.}~\bibnamefont
  {Nojiri}}, \bibinfo {author} {\bibfnamefont {S.~D.}\ \bibnamefont
  {Odintsov}}, \ and\ \bibinfo {author} {\bibfnamefont {V.~K.}\ \bibnamefont
  {Oikonomou}},\ }\href {\doibase 10.1016/j.physrep.2017.06.001} {\bibfield
  {journal} {\bibinfo  {journal} {Phys. Rept.}\ }\textbf {\bibinfo {volume}
  {692}},\ \bibinfo {pages} {1} (\bibinfo {year} {2017})},\ \Eprint
  {http://arxiv.org/abs/1705.11098} {arXiv:1705.11098 [gr-qc]} \BibitemShut
  {NoStop}%
\bibitem [{\citenamefont {Bahamonde}\ \emph {et~al.}(2023)\citenamefont
  {Bahamonde}, \citenamefont {Dialektopoulos}, \citenamefont
  {Escamilla-Rivera}, \citenamefont {Farrugia}, \citenamefont {Gakis},
  \citenamefont {Hendry}, \citenamefont {Hohmann}, \citenamefont {Levi~Said},
  \citenamefont {Mifsud},\ and\ \citenamefont {Di~Valentino}}]{Celia}%
  \BibitemOpen
  \bibfield  {author} {\bibinfo {author} {\bibfnamefont {S.}~\bibnamefont
  {Bahamonde}}, \bibinfo {author} {\bibfnamefont {K.~F.}\ \bibnamefont
  {Dialektopoulos}}, \bibinfo {author} {\bibfnamefont {C.}~\bibnamefont
  {Escamilla-Rivera}}, \bibinfo {author} {\bibfnamefont {G.}~\bibnamefont
  {Farrugia}}, \bibinfo {author} {\bibfnamefont {V.}~\bibnamefont {Gakis}},
  \bibinfo {author} {\bibfnamefont {M.}~\bibnamefont {Hendry}}, \bibinfo
  {author} {\bibfnamefont {M.}~\bibnamefont {Hohmann}}, \bibinfo {author}
  {\bibfnamefont {J.}~\bibnamefont {Levi~Said}}, \bibinfo {author}
  {\bibfnamefont {J.}~\bibnamefont {Mifsud}}, \ and\ \bibinfo {author}
  {\bibfnamefont {E.}~\bibnamefont {Di~Valentino}},\ }\href {\doibase
  10.1088/1361-6633/ac9cef} {\bibfield  {journal} {\bibinfo  {journal} {Rept.
  Prog. Phys.}\ }\textbf {\bibinfo {volume} {86}},\ \bibinfo {pages} {026901}
  (\bibinfo {year} {2023})},\ \Eprint {http://arxiv.org/abs/2106.13793}
  {arXiv:2106.13793 [gr-qc]} \BibitemShut {NoStop}%
\bibitem [{\citenamefont {Franco}\ \emph {et~al.}(2020)\citenamefont {Franco},
  \citenamefont {Escamilla-Rivera},\ and\ \citenamefont {Levi~Said}}]{Celia2}%
  \BibitemOpen
  \bibfield  {author} {\bibinfo {author} {\bibfnamefont {G.~A.~R.}\
  \bibnamefont {Franco}}, \bibinfo {author} {\bibfnamefont {C.}~\bibnamefont
  {Escamilla-Rivera}}, \ and\ \bibinfo {author} {\bibfnamefont
  {J.}~\bibnamefont {Levi~Said}},\ }\href {\doibase
  10.1140/epjc/s10052-020-8253-7} {\bibfield  {journal} {\bibinfo  {journal}
  {Eur. Phys. J. C}\ }\textbf {\bibinfo {volume} {80}},\ \bibinfo {pages} {677}
  (\bibinfo {year} {2020})},\ \Eprint {http://arxiv.org/abs/2005.14191}
  {arXiv:2005.14191 [gr-qc]} \BibitemShut {NoStop}%
\bibitem [{\citenamefont {Heisenberg}\ \emph {et~al.}(2023)\citenamefont
  {Heisenberg}, \citenamefont {Hohmann},\ and\ \citenamefont {Kuhn}}]{Lavinia}%
  \BibitemOpen
  \bibfield  {author} {\bibinfo {author} {\bibfnamefont {L.}~\bibnamefont
  {Heisenberg}}, \bibinfo {author} {\bibfnamefont {M.}~\bibnamefont {Hohmann}},
  \ and\ \bibinfo {author} {\bibfnamefont {S.}~\bibnamefont {Kuhn}},\ }\href
  {\doibase 10.1140/epjc/s10052-023-11462-6} {\bibfield  {journal} {\bibinfo
  {journal} {Eur. Phys. J. C}\ }\textbf {\bibinfo {volume} {83}},\ \bibinfo
  {pages} {315} (\bibinfo {year} {2023})},\ \Eprint
  {http://arxiv.org/abs/2212.14324} {arXiv:2212.14324 [gr-qc]} \BibitemShut
  {NoStop}%
\bibitem [{\citenamefont {Akrami}\ \emph {et~al.}(2021)\citenamefont {Akrami}
  \emph {et~al.}}]{CANTATA}%
  \BibitemOpen
  \bibfield  {author} {\bibinfo {author} {\bibfnamefont {Y.}~\bibnamefont
  {Akrami}} \emph {et~al.} (\bibinfo {collaboration} {CANTATA}),\ }\href
  {\doibase 10.1007/978-3-030-83715-0} {\emph {\bibinfo {title} {{Modified
  Gravity and Cosmology}: {An Update by the CANTATA Network}}}},\ edited by\
  \bibinfo {editor} {\bibfnamefont {E.~N.}\ \bibnamefont {Saridakis}}, \bibinfo
  {editor} {\bibfnamefont {R.}~\bibnamefont {Lazkoz}}, \bibinfo {editor}
  {\bibfnamefont {V.}~\bibnamefont {Salzano}}, \bibinfo {editor} {\bibfnamefont
  {P.}~\bibnamefont {Vargas~Moniz}}, \bibinfo {editor} {\bibfnamefont
  {S.}~\bibnamefont {Capozziello}}, \bibinfo {editor} {\bibfnamefont
  {J.}~\bibnamefont {Beltr\'an~Jim\'enez}}, \bibinfo {editor} {\bibfnamefont
  {M.}~\bibnamefont {De~Laurentis}}, \ and\ \bibinfo {editor} {\bibfnamefont
  {G.~J.}\ \bibnamefont {Olmo}}\ (\bibinfo  {publisher} {Springer},\ \bibinfo
  {year} {2021})\ \Eprint {http://arxiv.org/abs/2105.12582} {arXiv:2105.12582
  [gr-qc]} \BibitemShut {NoStop}%
\bibitem [{\citenamefont {{B{\"o}hmer}}\ and\ \citenamefont
  {{Jensko}}(2021)}]{Bohmer2021}%
  \BibitemOpen
  \bibfield  {author} {\bibinfo {author} {\bibfnamefont {C.~G.}\ \bibnamefont
  {{B{\"o}hmer}}}\ and\ \bibinfo {author} {\bibfnamefont {E.}~\bibnamefont
  {{Jensko}}},\ }\href {\doibase 10.1103/PhysRevD.104.024010} {\bibfield
  {journal} {\bibinfo  {journal} {Phys. Rev. D}\ }\textbf {\bibinfo {volume}
  {104}},\ \bibinfo {eid} {024010} (\bibinfo {year} {2021})},\ \Eprint
  {http://arxiv.org/abs/2103.15906} {arXiv:2103.15906 [gr-qc]} \BibitemShut
  {NoStop}%
\bibitem [{\citenamefont {{B{\"o}hmer}}\ and\ \citenamefont
  {{Jensko}}(2023)}]{Bohmer2023}%
  \BibitemOpen
  \bibfield  {author} {\bibinfo {author} {\bibfnamefont {C.~G.}\ \bibnamefont
  {{B{\"o}hmer}}}\ and\ \bibinfo {author} {\bibfnamefont {E.}~\bibnamefont
  {{Jensko}}},\ }\href {\doibase 10.1063/5.0150038} {\bibfield  {journal}
  {\bibinfo  {journal} {Journal of Mathematical Physics}\ }\textbf {\bibinfo
  {volume} {64}},\ \bibinfo {eid} {082505} (\bibinfo {year} {2023})},\ \Eprint
  {http://arxiv.org/abs/2301.11051} {arXiv:2301.11051 [gr-qc]} \BibitemShut
  {NoStop}%
\bibitem [{\citenamefont {{Beltr{\'a}n Jim{\'e}nez}}\ \emph
  {et~al.}(2018)\citenamefont {{Beltr{\'a}n Jim{\'e}nez}}, \citenamefont
  {{Heisenberg}},\ and\ \citenamefont {{Koivisto}}}]{Beltran2018}%
  \BibitemOpen
  \bibfield  {author} {\bibinfo {author} {\bibfnamefont {J.}~\bibnamefont
  {{Beltr{\'a}n Jim{\'e}nez}}}, \bibinfo {author} {\bibfnamefont
  {L.}~\bibnamefont {{Heisenberg}}}, \ and\ \bibinfo {author} {\bibfnamefont
  {T.~S.}\ \bibnamefont {{Koivisto}}},\ }\href {\doibase
  10.1088/1475-7516/2018/08/039} {\bibfield  {journal} {\bibinfo  {journal}
  {JCAP}\ }\textbf {\bibinfo {volume} {2018}},\ \bibinfo {eid} {039} (\bibinfo
  {year} {2018})},\ \Eprint {http://arxiv.org/abs/1803.10185} {arXiv:1803.10185
  [gr-qc]} \BibitemShut {NoStop}%
\bibitem [{\citenamefont {{D'Ambrosio}}\ \emph {et~al.}(2022)\citenamefont
  {{D'Ambrosio}}, \citenamefont {{Fell}}, \citenamefont {{Heisenberg}},\ and\
  \citenamefont {{Kuhn}}}]{DAmbrosio2022}%
  \BibitemOpen
  \bibfield  {author} {\bibinfo {author} {\bibfnamefont {F.}~\bibnamefont
  {{D'Ambrosio}}}, \bibinfo {author} {\bibfnamefont {S.~D.~B.}\ \bibnamefont
  {{Fell}}}, \bibinfo {author} {\bibfnamefont {L.}~\bibnamefont
  {{Heisenberg}}}, \ and\ \bibinfo {author} {\bibfnamefont {S.}~\bibnamefont
  {{Kuhn}}},\ }\href {\doibase 10.1103/PhysRevD.105.024042} {\bibfield
  {journal} {\bibinfo  {journal} {Phys. Rev. D}\ }\textbf {\bibinfo {volume}
  {105}},\ \bibinfo {eid} {024042} (\bibinfo {year} {2022})},\ \Eprint
  {http://arxiv.org/abs/2109.03174} {arXiv:2109.03174 [gr-qc]} \BibitemShut
  {NoStop}%
\bibitem [{\citenamefont {{Bajardi}}\ \emph {et~al.}(2020)\citenamefont
  {{Bajardi}}, \citenamefont {{Vernieri}},\ and\ \citenamefont
  {{Capozziello}}}]{Bajardi2020}%
  \BibitemOpen
  \bibfield  {author} {\bibinfo {author} {\bibfnamefont {F.}~\bibnamefont
  {{Bajardi}}}, \bibinfo {author} {\bibfnamefont {D.}~\bibnamefont
  {{Vernieri}}}, \ and\ \bibinfo {author} {\bibfnamefont {S.}~\bibnamefont
  {{Capozziello}}},\ }\href {\doibase 10.1140/epjp/s13360-020-00918-3}
  {\bibfield  {journal} {\bibinfo  {journal} {European Physical Journal Plus}\
  }\textbf {\bibinfo {volume} {135}},\ \bibinfo {eid} {912} (\bibinfo {year}
  {2020})},\ \Eprint {http://arxiv.org/abs/2011.01248} {arXiv:2011.01248
  [gr-qc]} \BibitemShut {NoStop}%
\bibitem [{\citenamefont {{Capozziello}}\ and\ \citenamefont
  {{D'Agostino}}(2022)}]{Rocco2022}%
  \BibitemOpen
  \bibfield  {author} {\bibinfo {author} {\bibfnamefont {S.}~\bibnamefont
  {{Capozziello}}}\ and\ \bibinfo {author} {\bibfnamefont {R.}~\bibnamefont
  {{D'Agostino}}},\ }\href {\doibase 10.1016/j.physletb.2022.137229} {\bibfield
   {journal} {\bibinfo  {journal} {Physics Letters B}\ }\textbf {\bibinfo
  {volume} {832}},\ \bibinfo {eid} {137229} (\bibinfo {year} {2022})},\ \Eprint
  {http://arxiv.org/abs/2204.01015} {arXiv:2204.01015 [gr-qc]} \BibitemShut
  {NoStop}%
\bibitem [{\citenamefont {{Solanki}}\ \emph {et~al.}(2022)\citenamefont
  {{Solanki}}, \citenamefont {{De}}, \citenamefont {{Mandal}},\ and\
  \citenamefont {{Sahoo}}}]{Solanki2022}%
  \BibitemOpen
  \bibfield  {author} {\bibinfo {author} {\bibfnamefont {R.}~\bibnamefont
  {{Solanki}}}, \bibinfo {author} {\bibfnamefont {A.}~\bibnamefont {{De}}},
  \bibinfo {author} {\bibfnamefont {S.}~\bibnamefont {{Mandal}}}, \ and\
  \bibinfo {author} {\bibfnamefont {P.~K.}\ \bibnamefont {{Sahoo}}},\ }\href
  {\doibase 10.1016/j.dark.2022.101053} {\bibfield  {journal} {\bibinfo
  {journal} {Physics of the Dark Universe}\ }\textbf {\bibinfo {volume} {36}},\
  \bibinfo {eid} {101053} (\bibinfo {year} {2022})},\ \Eprint
  {http://arxiv.org/abs/2201.06521} {arXiv:2201.06521 [gr-qc]} \BibitemShut
  {NoStop}%
\bibitem [{\citenamefont {Capozziello}\ and\ \citenamefont
  {Shokri}(2022)}]{Shokri}%
  \BibitemOpen
  \bibfield  {author} {\bibinfo {author} {\bibfnamefont {S.}~\bibnamefont
  {Capozziello}}\ and\ \bibinfo {author} {\bibfnamefont {M.}~\bibnamefont
  {Shokri}},\ }\href {\doibase 10.1016/j.dark.2022.101113} {\bibfield
  {journal} {\bibinfo  {journal} {Phys. Dark Univ.}\ }\textbf {\bibinfo
  {volume} {37}},\ \bibinfo {pages} {101113} (\bibinfo {year} {2022})},\
  \Eprint {http://arxiv.org/abs/2209.06670} {arXiv:2209.06670 [gr-qc]}
  \BibitemShut {NoStop}%
\bibitem [{\citenamefont {{Soudi}}\ \emph {et~al.}(2019)\citenamefont
  {{Soudi}}, \citenamefont {{Farrugia}}, \citenamefont {{Said}}, \citenamefont
  {{Gakis}},\ and\ \citenamefont {{Saridakis}}}]{Soudi2019}%
  \BibitemOpen
  \bibfield  {author} {\bibinfo {author} {\bibfnamefont {I.}~\bibnamefont
  {{Soudi}}}, \bibinfo {author} {\bibfnamefont {G.}~\bibnamefont {{Farrugia}}},
  \bibinfo {author} {\bibfnamefont {J.~L.}\ \bibnamefont {{Said}}}, \bibinfo
  {author} {\bibfnamefont {V.}~\bibnamefont {{Gakis}}}, \ and\ \bibinfo
  {author} {\bibfnamefont {E.~N.}\ \bibnamefont {{Saridakis}}},\ }\href
  {\doibase 10.1103/PhysRevD.100.044008} {\bibfield  {journal} {\bibinfo
  {journal} {Phys. Rev. D}\ }\textbf {\bibinfo {volume} {100}},\ \bibinfo {eid}
  {044008} (\bibinfo {year} {2019})},\ \Eprint
  {http://arxiv.org/abs/1810.08220} {arXiv:1810.08220 [gr-qc]} \BibitemShut
  {NoStop}%
\bibitem [{\citenamefont {D'Agostino}\ and\ \citenamefont
  {Nunes}(2022)}]{Dagostino2022}%
  \BibitemOpen
  \bibfield  {author} {\bibinfo {author} {\bibfnamefont {R.}~\bibnamefont
  {D'Agostino}}\ and\ \bibinfo {author} {\bibfnamefont {R.~C.}\ \bibnamefont
  {Nunes}},\ }\href {\doibase 10.1103/PhysRevD.106.124053} {\bibfield
  {journal} {\bibinfo  {journal} {Phys. Rev. D}\ }\textbf {\bibinfo {volume}
  {106}},\ \bibinfo {pages} {124053} (\bibinfo {year} {2022})}\BibitemShut
  {NoStop}%
\bibitem [{\citenamefont {Javed}\ \emph {et~al.}(2023)\citenamefont {Javed},
  \citenamefont {Mustafa}, \citenamefont {Mumtaz},\ and\ \citenamefont
  {Atamurotov}}]{Mustafa}%
  \BibitemOpen
  \bibfield  {author} {\bibinfo {author} {\bibfnamefont {F.}~\bibnamefont
  {Javed}}, \bibinfo {author} {\bibfnamefont {G.}~\bibnamefont {Mustafa}},
  \bibinfo {author} {\bibfnamefont {S.}~\bibnamefont {Mumtaz}}, \ and\ \bibinfo
  {author} {\bibfnamefont {F.}~\bibnamefont {Atamurotov}},\ }\href {\doibase
  10.1016/j.nuclphysb.2023.116180} {\bibfield  {journal} {\bibinfo  {journal}
  {Nucl. Phys. B}\ }\textbf {\bibinfo {volume} {990}},\ \bibinfo {pages}
  {116180} (\bibinfo {year} {2023})}\BibitemShut {NoStop}%
\bibitem [{\citenamefont {Parsaei}\ \emph {et~al.}(2022)\citenamefont
  {Parsaei}, \citenamefont {Rastgoo},\ and\ \citenamefont
  {Sahoo}}]{Parsaei2022}%
  \BibitemOpen
  \bibfield  {author} {\bibinfo {author} {\bibfnamefont {F.}~\bibnamefont
  {Parsaei}}, \bibinfo {author} {\bibfnamefont {S.}~\bibnamefont {Rastgoo}}, \
  and\ \bibinfo {author} {\bibfnamefont {P.~K.}\ \bibnamefont {Sahoo}},\ }\href
  {\doibase 10.1140/epjp/s13360-022-03298-y} {\bibfield  {journal} {\bibinfo
  {journal} {Eur. Phys. J. Plus}\ }\textbf {\bibinfo {volume} {137}},\ \bibinfo
  {pages} {1083} (\bibinfo {year} {2022})},\ \Eprint
  {http://arxiv.org/abs/2203.06374} {arXiv:2203.06374 [gr-qc]} \BibitemShut
  {NoStop}%
\bibitem [{\citenamefont {Sharma}\ \emph {et~al.}(2022)\citenamefont {Sharma},
  \citenamefont {Shweta},\ and\ \citenamefont {Mishra}}]{Sharma}%
  \BibitemOpen
  \bibfield  {author} {\bibinfo {author} {\bibfnamefont {U.~K.}\ \bibnamefont
  {Sharma}}, \bibinfo {author} {\bibnamefont {Shweta}}, \ and\ \bibinfo
  {author} {\bibfnamefont {A.~K.}\ \bibnamefont {Mishra}},\ }\href {\doibase
  10.1142/S0219887822500190} {\bibfield  {journal} {\bibinfo  {journal} {Int.
  J. Geom. Meth. Mod. Phys.}\ }\textbf {\bibinfo {volume} {19}},\ \bibinfo
  {pages} {2250019} (\bibinfo {year} {2022})},\ \Eprint
  {http://arxiv.org/abs/2108.07174} {arXiv:2108.07174 [physics.gen-ph]}
  \BibitemShut {NoStop}%
\bibitem [{\citenamefont {Dialektopoulos}\ \emph {et~al.}(2019)\citenamefont
  {Dialektopoulos}, \citenamefont {Koivisto},\ and\ \citenamefont
  {Capozziello}}]{Kostas}%
  \BibitemOpen
  \bibfield  {author} {\bibinfo {author} {\bibfnamefont {K.~F.}\ \bibnamefont
  {Dialektopoulos}}, \bibinfo {author} {\bibfnamefont {T.~S.}\ \bibnamefont
  {Koivisto}}, \ and\ \bibinfo {author} {\bibfnamefont {S.}~\bibnamefont
  {Capozziello}},\ }\href {\doibase 10.1140/epjc/s10052-019-7106-8} {\bibfield
  {journal} {\bibinfo  {journal} {Eur. Phys. J. C}\ }\textbf {\bibinfo {volume}
  {79}},\ \bibinfo {pages} {606} (\bibinfo {year} {2019})},\ \Eprint
  {http://arxiv.org/abs/1905.09019} {arXiv:1905.09019 [gr-qc]} \BibitemShut
  {NoStop}%
\bibitem [{\citenamefont {Bajardi}\ and\ \citenamefont
  {Capozziello}(2023)}]{Quantum}%
  \BibitemOpen
  \bibfield  {author} {\bibinfo {author} {\bibfnamefont {F.}~\bibnamefont
  {Bajardi}}\ and\ \bibinfo {author} {\bibfnamefont {S.}~\bibnamefont
  {Capozziello}},\ }\href {\doibase 10.1140/epjc/s10052-023-11703-8} {\bibfield
   {journal} {\bibinfo  {journal} {Eur. Phys. J. C}\ }\textbf {\bibinfo
  {volume} {83}},\ \bibinfo {pages} {531} (\bibinfo {year} {2023})},\ \Eprint
  {http://arxiv.org/abs/2305.00318} {arXiv:2305.00318 [gr-qc]} \BibitemShut
  {NoStop}%
\bibitem [{\citenamefont {Capozziello}\ \emph {et~al.}(2022)\citenamefont
  {Capozziello}, \citenamefont {De~Falco},\ and\ \citenamefont
  {Ferrara}}]{Capozziello2022}%
  \BibitemOpen
  \bibfield  {author} {\bibinfo {author} {\bibfnamefont {S.}~\bibnamefont
  {Capozziello}}, \bibinfo {author} {\bibfnamefont {V.}~\bibnamefont
  {De~Falco}}, \ and\ \bibinfo {author} {\bibfnamefont {C.}~\bibnamefont
  {Ferrara}},\ }\href {\doibase 10.1140/epjc/s10052-022-10823-x} {\bibfield
  {journal} {\bibinfo  {journal} {Eur. Phys. J. C}\ }\textbf {\bibinfo {volume}
  {82}},\ \bibinfo {pages} {865} (\bibinfo {year} {2022})},\ \Eprint
  {http://arxiv.org/abs/2208.03011} {arXiv:2208.03011 [gr-qc]} \BibitemShut
  {NoStop}%
\bibitem [{\citenamefont {Aldrovandi}\ and\ \citenamefont
  {Pereira}(2013)}]{Pereira}%
  \BibitemOpen
  \bibfield  {author} {\bibinfo {author} {\bibfnamefont {R.}~\bibnamefont
  {Aldrovandi}}\ and\ \bibinfo {author} {\bibfnamefont {J.~G.}\ \bibnamefont
  {Pereira}},\ }\href {\doibase 10.1007/978-94-007-5143-9} {\emph {\bibinfo
  {title} {{Teleparallel Gravity}: {An Introduction}}}}\ (\bibinfo  {publisher}
  {Springer},\ \bibinfo {year} {2013})\BibitemShut {NoStop}%
\bibitem [{\citenamefont {Cai}\ \emph {et~al.}(2016)\citenamefont {Cai},
  \citenamefont {Capozziello}, \citenamefont {De~Laurentis},\ and\
  \citenamefont {Saridakis}}]{Cai}%
  \BibitemOpen
  \bibfield  {author} {\bibinfo {author} {\bibfnamefont {Y.-F.}\ \bibnamefont
  {Cai}}, \bibinfo {author} {\bibfnamefont {S.}~\bibnamefont {Capozziello}},
  \bibinfo {author} {\bibfnamefont {M.}~\bibnamefont {De~Laurentis}}, \ and\
  \bibinfo {author} {\bibfnamefont {E.~N.}\ \bibnamefont {Saridakis}},\ }\href
  {\doibase 10.1088/0034-4885/79/10/106901} {\bibfield  {journal} {\bibinfo
  {journal} {Rept. Prog. Phys.}\ }\textbf {\bibinfo {volume} {79}},\ \bibinfo
  {pages} {106901} (\bibinfo {year} {2016})},\ \Eprint
  {http://arxiv.org/abs/1511.07586} {arXiv:1511.07586 [gr-qc]} \BibitemShut
  {NoStop}%
\bibitem [{\citenamefont {Beltr\'an~Jim\'enez}\ \emph
  {et~al.}(2019)\citenamefont {Beltr\'an~Jim\'enez}, \citenamefont
  {Heisenberg},\ and\ \citenamefont {Koivisto}}]{Tomi}%
  \BibitemOpen
  \bibfield  {author} {\bibinfo {author} {\bibfnamefont {J.}~\bibnamefont
  {Beltr\'an~Jim\'enez}}, \bibinfo {author} {\bibfnamefont {L.}~\bibnamefont
  {Heisenberg}}, \ and\ \bibinfo {author} {\bibfnamefont {T.~S.}\ \bibnamefont
  {Koivisto}},\ }\href {\doibase 10.3390/universe5070173} {\bibfield  {journal}
  {\bibinfo  {journal} {Universe}\ }\textbf {\bibinfo {volume} {5}},\ \bibinfo
  {pages} {173} (\bibinfo {year} {2019})},\ \Eprint
  {http://arxiv.org/abs/1903.06830} {arXiv:1903.06830 [hep-th]} \BibitemShut
  {NoStop}%
\bibitem [{\citenamefont {{Bahamonde}}\ \emph {et~al.}(2015)\citenamefont
  {{Bahamonde}}, \citenamefont {{B{\"o}hmer}},\ and\ \citenamefont
  {{Wright}}}]{Bahamonde2015}%
  \BibitemOpen
  \bibfield  {author} {\bibinfo {author} {\bibfnamefont {S.}~\bibnamefont
  {{Bahamonde}}}, \bibinfo {author} {\bibfnamefont {C.~G.}\ \bibnamefont
  {{B{\"o}hmer}}}, \ and\ \bibinfo {author} {\bibfnamefont {M.}~\bibnamefont
  {{Wright}}},\ }\href {\doibase 10.1103/PhysRevD.92.104042} {\bibfield
  {journal} {\bibinfo  {journal} {Phys. Rev. D}\ }\textbf {\bibinfo {volume} {92}},\
  \bibinfo {eid} {104042} (\bibinfo {year} {2015})},\ \Eprint
  {http://arxiv.org/abs/1508.05120} {arXiv:1508.05120 [gr-qc]} \BibitemShut
  {NoStop}%
\bibitem [{\citenamefont {Bahamonde}\ and\ \citenamefont
  {Capozziello}(2017)}]{Sebastian}%
  \BibitemOpen
  \bibfield  {author} {\bibinfo {author} {\bibfnamefont {S.}~\bibnamefont
  {Bahamonde}}\ and\ \bibinfo {author} {\bibfnamefont {S.}~\bibnamefont
  {Capozziello}},\ }\href {\doibase 10.1140/epjc/s10052-017-4677-0} {\bibfield
  {journal} {\bibinfo  {journal} {Eur. Phys. J. C}\ }\textbf {\bibinfo {volume}
  {77}},\ \bibinfo {pages} {107} (\bibinfo {year} {2017})},\ \Eprint
  {http://arxiv.org/abs/1612.01299} {arXiv:1612.01299 [gr-qc]} \BibitemShut
  {NoStop}%
\bibitem [{\citenamefont {Capozziello}\ \emph
  {et~al.}(2020{\natexlab{a}})\citenamefont {Capozziello}, \citenamefont
  {Capriolo},\ and\ \citenamefont {Caso}}]{Capriolo}%
  \BibitemOpen
  \bibfield  {author} {\bibinfo {author} {\bibfnamefont {S.}~\bibnamefont
  {Capozziello}}, \bibinfo {author} {\bibfnamefont {M.}~\bibnamefont
  {Capriolo}}, \ and\ \bibinfo {author} {\bibfnamefont {L.}~\bibnamefont
  {Caso}},\ }\href {\doibase 10.1140/epjc/s10052-020-7737-9} {\bibfield
  {journal} {\bibinfo  {journal} {Eur. Phys. J. C}\ }\textbf {\bibinfo {volume}
  {80}},\ \bibinfo {pages} {156} (\bibinfo {year} {2020}{\natexlab{a}})},\
  \Eprint {http://arxiv.org/abs/1912.12469} {arXiv:1912.12469 [gr-qc]}
  \BibitemShut {NoStop}%
\bibitem [{\citenamefont {Gibbons}\ and\ \citenamefont
  {Hawking}(1977)}]{Gibbons1977}%
  \BibitemOpen
  \bibfield  {author} {\bibinfo {author} {\bibfnamefont {G.~W.}\ \bibnamefont
  {Gibbons}}\ and\ \bibinfo {author} {\bibfnamefont {S.~W.}\ \bibnamefont
  {Hawking}},\ }\href {\doibase 10.1103/PhysRevD.15.2752} {\bibfield  {journal}
  {\bibinfo  {journal} {Phys. Rev. D}\ }\textbf {\bibinfo {volume} {15}},\
  \bibinfo {pages} {2752} (\bibinfo {year} {1977})}\BibitemShut {NoStop}%
\bibitem [{\citenamefont {Romano}\ and\ \citenamefont
  {Mango~Furnari}(2019)}]{Romano2019}%
  \BibitemOpen
  \bibfield  {author} {\bibinfo {author} {\bibfnamefont {A.}~\bibnamefont
  {Romano}}\ and\ \bibinfo {author} {\bibfnamefont {M.}~\bibnamefont
  {Mango~Furnari}},\ }\href@noop {} {\emph {\bibinfo {title} {The Physical and
  Mathematical Foundations of the Theory of Relativity: A Critical Analysis}}}\
  (\bibinfo  {publisher} {Springer International Publishing},\ \bibinfo
  {address} {Cham},\ \bibinfo {year} {2019})\BibitemShut {NoStop}%
\bibitem [{\citenamefont {Dyer}\ and\ \citenamefont
  {Hinterbichler}(2009)}]{Dyer2009}%
  \BibitemOpen
  \bibfield  {author} {\bibinfo {author} {\bibfnamefont {E.}~\bibnamefont
  {Dyer}}\ and\ \bibinfo {author} {\bibfnamefont {K.}~\bibnamefont
  {Hinterbichler}},\ }\href {\doibase 10.1103/PhysRevD.79.024028} {\bibfield
  {journal} {\bibinfo  {journal} {Phys. Rev. D}\ }\textbf {\bibinfo {volume}
  {79}},\ \bibinfo {pages} {024028} (\bibinfo {year} {2009})}\BibitemShut
  {NoStop}%
\bibitem [{\citenamefont {Capozziello}\ \emph
  {et~al.}(2020{\natexlab{b}})\citenamefont {Capozziello}, \citenamefont
  {Capriolo},\ and\ \citenamefont {Caso}}]{Capriolo2}%
  \BibitemOpen
  \bibfield  {author} {\bibinfo {author} {\bibfnamefont {S.}~\bibnamefont
  {Capozziello}}, \bibinfo {author} {\bibfnamefont {M.}~\bibnamefont
  {Capriolo}}, \ and\ \bibinfo {author} {\bibfnamefont {L.}~\bibnamefont
  {Caso}},\ }\href {\doibase 10.1088/1361-6382/abbe71} {\bibfield  {journal}
  {\bibinfo  {journal} {Class. Quant. Grav.}\ }\textbf {\bibinfo {volume}
  {37}},\ \bibinfo {pages} {235013} (\bibinfo {year} {2020}{\natexlab{b}})},\
  \Eprint {http://arxiv.org/abs/2010.00451} {arXiv:2010.00451 [gr-qc]}
  \BibitemShut {NoStop}%
\bibitem [{\citenamefont {Starobinsky}(1980)}]{Starobinsky}%
  \BibitemOpen
  \bibfield  {author} {\bibinfo {author} {\bibfnamefont {A.~A.}\ \bibnamefont
  {Starobinsky}},\ }\href {\doibase 10.1016/0370-2693(80)90670-X} {\bibfield
  {journal} {\bibinfo  {journal} {Phys. Lett. B}\ }\textbf {\bibinfo {volume}
  {91}},\ \bibinfo {pages} {99} (\bibinfo {year} {1980})}\BibitemShut {NoStop}%
\bibitem [{\citenamefont {Magnus}\ and\ \citenamefont
  {Neudecker}(1999)}]{Magnus1999}%
  \BibitemOpen
  \bibfield  {author} {\bibinfo {author} {\bibfnamefont {J.~R.}\ \bibnamefont
  {Magnus}}\ and\ \bibinfo {author} {\bibfnamefont {H.}~\bibnamefont
  {Neudecker}},\ }\href@noop {} {\emph {\bibinfo {title} {Matrix Differential
  Calculus with Applications in Statistics and Econometrics}}},\ \bibinfo
  {edition} {2nd}\ ed.\ (\bibinfo  {publisher} {John Wiley},\ \bibinfo {year}
  {1999})\BibitemShut {NoStop}%
\bibitem [{\citenamefont {{Xu}}\ \emph {et~al.}(2019)\citenamefont {{Xu}},
  \citenamefont {{Li}}, \citenamefont {{Harko}},\ and\ \citenamefont
  {{Liang}}}]{Xu2019}%
  \BibitemOpen
  \bibfield  {author} {\bibinfo {author} {\bibfnamefont {Y.}~\bibnamefont
  {{Xu}}}, \bibinfo {author} {\bibfnamefont {G.}~\bibnamefont {{Li}}}, \bibinfo
  {author} {\bibfnamefont {T.}~\bibnamefont {{Harko}}}, \ and\ \bibinfo
  {author} {\bibfnamefont {S.-D.}\ \bibnamefont {{Liang}}},\ }\href {\doibase
  10.1140/epjc/s10052-019-7207-4} {\bibfield  {journal} {\bibinfo  {journal}
  {European Physical Journal C}\ }\textbf {\bibinfo {volume} {79}},\ \bibinfo
  {eid} {708} (\bibinfo {year} {2019})},\ \Eprint
  {http://arxiv.org/abs/1908.04760} {arXiv:1908.04760 [gr-qc]} \BibitemShut
  {NoStop}%
\end{thebibliography}
\end{document}